# Layer dependent antiferromagnetism in the Sr$_4$Ru$_3$O$_{10}$ ruthenate at the metamagnetic-like transition

L. Capogna[1], V. Granata[2], B. Ouladdiaf[3], J. A. Rodriguez-Velamazan[3], R. Fittipaldi[2] and A. Vecchione[2].

[1]*Consiglio Nazionale delle Ricerche, Institut Laue Langevin, Av. Des Martyrs, 38040 Grenoble - France*

[2] *Dipartimento di Fisica 'E R Caianiello', Università degli Studi di Salerno, Via Ponte don Melillo I-84084 Fisciano - Italy*

[3]*Institut Laue Langevin, Av. Des Martyrs, 38040 Grenoble – France*

## Abstract

We have investigated the metamagnetic-like transition in the triple layer ruthenate Sr$_4$Ru$_3$O$_{10}$ by means of neutron diffraction from single crystals. The magnetic structure of the compound appears to be determined in a complex way by the two substructures of inequivalent ruthenium ions. At T$_c$=105K the system has a sharp transition into a ferromagnetic state along the *c*-axis which is driven by the ruthenium atoms in the central octahedra of the triple layers whereas the substructure of the outer ruthenium atoms tend to align in the *ab* plane achieving an antiferromagnetic order at the metamagnetic transition T*~50K. Below T* the strong anisotropy along *c* prevails, the outer ruthenium tend to align along the *c*-axis and the in-plane antiferromagnetic order disappears. This finding confirms the delicate balance between antiferro and ferromagnetic couplings in the (Sr,Ca)$_{n+1}$Ru$_n$O$_{3n+1}$ family of compounds, and proves the layer dependence of the magnetic anisotropy in Sr$_4$Ru$_3$O$_{10}$.

1. ## Introduction

The triple layer Sr$_4$Ru$_3$O$_{10}$ belongs to the Ruddlesden-Popper perovskite ruthenates (Sr,Ca)$_{n+1}$Ru$_n$O$_{3n+1}$, a family of 4*d* transition-metal oxides whose magnetic and electronic properties are sensitively dependent on the layer number *n* and on the structural distortions.

For instance, the single-layer Sr$_2$RuO$_4$ (*n* = 1) shows an unconventional superconducting state [1], whereas the ground state of the double-layer Sr$_3$Ru$_2$O$_7$ (*n* = 2) is a Fermi liquid close to a ferromagnetic instability [2]. The three-dimensional SrRuO$_3$ (*n* = ∞) is a ferromagnetic metal with a Curie temperature Tc = 160 K [3,4]. On the other hand, the calcium ruthenate Ca$_2$RuO$_4$ (*n* = 1) is an antiferromagnetic Mott insulator with a Neel temperature T$_N$ ~ 110 K [5], while Ca$_3$Ru$_2$O$_7$ (*n* = 2) exhibits a quasi-two-dimensional metallic behavior and becomes antiferromagnetic below T$_N$ ~ 56 K [6,7]. Last, CaRuO$_3$ is a paramagnetic metal [8,9].

The triple-layer Sr$_4$Ru$_3$O$_{10}$ has been attracting considerable interest because of its complex structure and its unusual magnetic behaviour. Susceptibility measurements show a paramagnetic-ferromagnetic transition at Tc = 105 K, below which the easy axis is along the *c* direction [10]; a minor transition is observed at around T*=50K et it is referred to as the metamagnetic transition since below this temperature the magnetisation shows a sudden increase at a magnetic field of about 2 T. Curiously, while the metamagnetic transition is a well pronounced feature in the magnetisation and in the resistitity [10-14], it does not appear in the specific heat curves [15]. Although several distinct scenarios have been proposed to account for the anomaly at T*, its

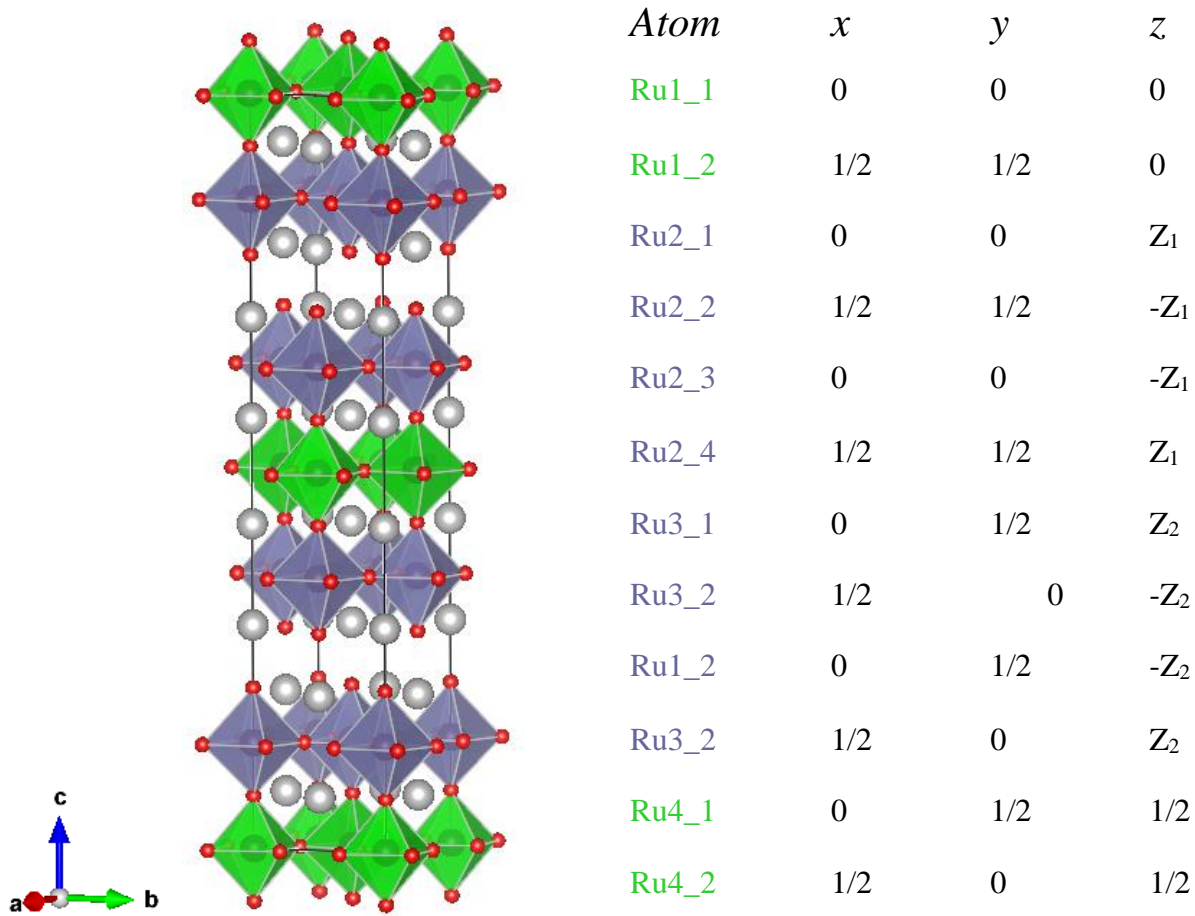

| Atom | x | y | z |
|---|---|---|---|
| Ru1_1 | 0 | 0 | 0 |
| Ru1_2 | 1/2 | 1/2 | 0 |
| Ru2_1 | 0 | 0 | $Z_1$ |
| Ru2_2 | 1/2 | 1/2 | $-Z_1$ |
| Ru2_3 | 0 | 0 | $-Z_1$ |
| Ru2_4 | 1/2 | 1/2 | $Z_1$ |
| Ru3_1 | 0 | 1/2 | $Z_2$ |
| Ru3_2 | 1/2 | 0 | $-Z_2$ |
| Ru1_2 | 0 | 1/2 | $-Z_2$ |
| Ru3_2 | 1/2 | 0 | $Z_2$ |
| Ru4_1 | 0 | 1/2 | 1/2 |
| Ru4_2 | 1/2 | 0 | 1/2 |

Figure1 Crystal structure of $Sr_4Ru_3O_{10}$ at room temperature [10]. The green octahedra are at the centre of the trilayers. The external blue octahedra are located at Z1 and Z2 along the *c*-axis. The atom positions are listed for the twelve ruthenium in the unit cell.

intrinsic character remains an open question. Many experimental techniques, including neutron scattering [16,17], Raman scattering [18] and resistivity [19] have been used to try to clarify the origin of T* and some contradictory pictures have been put forward. From Raman and resistitity experiments it has been inferred that the metamagnetism would arise from an antiferromagnetic (AF) component in the *ab* plane due to canting of all spins; conversely initial neutron diffraction experiments discarded [16] or overlooked [17] the possibility of an AF component. In our previous neutron study for instance, we focussed on the low temperature magnetic structure (T=1.5K) which was found to have a magnetic propagation vector K=(0,0,0) with all magnetic moments ferromagnetically aligned along *c* [17]. The clear anomaly at T*=50 K that we observed in the temperature dependence of the Bragg peaks [17] was not interpreted in terms of an additional magnetic phase due to the absence of convincing magnetic signal in the *ab* plane within our experimental conditions. More recently, the hyphothesis of a ferromagnetic component in the *ab*-plane at around T* has been put forward by Zhu and collaborators, based on the observation of a sizable neutron magnetic scattering in the *ab* plane [20]. However the proposed model does not fit entirely our previous experimental observations [17], especially the lack of intensity in some specific reflections. On the basis of symmetry analysis and a few additional neutron measurements, we suggest in this paper an alternative scenario in which the metamagnetic transition would stem from an antiferromagnetic ordering in the substructure of the ruthenium atoms in the outer octahedra of the triple layers, almost indipendently of the substructure of the inner ruthenium atoms which are predominantly ferromagnetic.

1. **Results and discussion**

The experiments were carried out at the Institut Laue Langevin in Grenoble on the crystal diffractometer D10 in normal beam configuration and equipped with a two-dimensional detector and a vertical cryomagnet. A wavelength of 2.36 Å was chosen in order to measure the magnetic reflections. The sample is the single crystal of dimensions ~$L_xL_yL_z$=3x2x0.5 mm$^3$ reported in Ref.17, it was grown with the floating zone technique [21] and was oriented with the [-h,h,0] direction along the vertical axis while a magnetic vertical field was varied between 0 and 6 tesla. The crystal structure has a primitive space group [10] with a rather long *c* axis~28 Å. The magnetically active part of the unit cell consists of four blocks of ruthenium atoms octahedrally coordinated with oxygens; two of such blocks are centred in the basal plane, at z=0 and two blocks are located at z=0.5, see Figure 1. As a result, each cell contains four formula units with a total of twelve ruthenium atoms in four different Wickoff sites: the ruthenium atoms at the centre of the octahedra (Ru_in) are in a more symmetric site with multeplicity 2, whilst the ruthenium atoms in the external octahedra of the trilayers (Ru_out) have multeplicity 4. At room temperature the inner octahedra are regular contrary to the outer ones which are slightly elongated [10], the inner octahedra are also rotated with an angle of rotation above the critical angle for ferromagnetism [21]. As a consequence, the Ru_in are supposed to be more prone to ferromagnetism than the Ru_out. This is reflected in the value of the ferromagnetic moment along the z axis on the two sites, which has been measured with neutron scattering at low temperature (1.5K) yielding 1.59 $\mu_B$ on the inner ruthenium atoms and 0.92 $\mu_B$ on the outer ones [17]. The magnetic structure at T=1.5K is a ferromagnetic state consisting of ferromagnetic modes $f_z$ and $F_z$ on the inner and outer ruthenium atoms respectively [17]. However, the magnetic structure at the metamagnetic temperature T*=50K is still not entirely clear. Given the diversity of the inner and outer ruthenium atoms, we can speculate that the different sites be responsible for the two observed transitions at $T_C$=105 K and at T*=50K. In a recent report [20], neutron scattering experiments have been able to ascertain the presence of a magnetic signal in the *ab* plane about T*, based on the observation of additional intensity in the 002, 006 and 008 reflections. The hyphothesis of a ferromagnetic axis sligtly tilted in the *ab*-plane at around T* has been put forward to explain the occurrence of magnetism in the *ab* plane, however no explanation has been provided for the absence of the 004 reflection. We believe that the lack of intensity in the 004 be an important key to solving the puzzle. If the magnetic coupling in the *ab* plane was of ferromagnetic nature, then the magnetic contribution to the Bragg reflections would be comparable for the 004 and for the 002 reflection since for ferromagnetic modes in the *ab* plane (irreducible representations $\Gamma_3$ and $\Gamma_4$, see Table T1 and T2) the magnetic structure factor would be modulated with $l$ according to the periodic form $\cos[l\pi(z_1+z_2)]*\cos[l\pi(z_1-z_2)]$. Our measurements i zero field (Figure 2) show however that this is not the case, on the contrary at around T* the reflection 002 is much more intense than the 004 which is in fact vanishing, and this difference is well beyond the attenuation of the magnetic scattering by the magnetic form factor at these two close scattering vectors [20].

Table T.1: Table of characters of the irreducible representations for $Sr_4Ru_3O_{10}$, space group P*bam* and magnetic propagation vector **k** = (0,0,0), calculated with BasIreps. The decomposition of the magnetic representation $\Gamma$mag in terms of the irreducible representations is given for the four ruthenium sites.

| Symm. oper. | {1\|000} | {2x\|pp0} | {2y\|pp0} | {2z\|000} | {-1\|000} | {myz\|pp0} | {mxz\|pp0} | {mxy\|000} |
|---|---|---|---|---|---|---|---|---|
| *IrRep* | | | | | | | | |
| $\Gamma_1$ | 1 | 1 | 1 | 1 | 1 | 1 | 1 | 1 |
| $\Gamma_2$ | 1 | -1 | -1 | 1 | 1 | 1 | -1 | 1 |
| $\Gamma_3$ | 1 | 1 | -1 | -1 | 1 | 1 | -1 | -1 |
| $\Gamma_4$ | 1 | -1 | 1 | -1 | 1 | 1 | 1 | -1 |
| $\Gamma_5$ | 1 | 1 | 1 | 1 | 1 | -1 | -1 | -1 |
| $\Gamma_6$ | 1 | -1 | -1 | 1 | 1 | -1 | 1 | -1 |
| $\Gamma_7$ | 1 | 1 | -1 | 1 | -1 | -1 | 1 | 1 |
| $\Gamma_8$ | 1 | -1 | 1 | 1 | -1 | -1 | -1 | 1 |
| $\Gamma$mag (Ru$_{1/4}$) | 1 $\Gamma_1$ + 1 $\Gamma_2$ + 2 $\Gamma_3$ + 2 $\Gamma_4$ | | | | | | | |
| $\Gamma$mag (Ru$_{2/3}$) | 1 $\Gamma_1$ + 1 $\Gamma_2$+ 2 $\Gamma_3$+ 2 $\Gamma_4$+ 1 $\Gamma_5$ + 1 $\Gamma_6$ + 2 $\Gamma_7$ + 2 $\Gamma_8$ | | | | | | | |

Table T2: Irreducible representations for P*bam* and **k** = (0,0,0,) and corresponding magnetic space groups. We have here adopted the convention for the modes proposed by Bertaut, *e.g.* J. Phys. Chem. Solids **28,** 2143 (1967): a(+ -) and f(++) for Ru1 and Ru4, F(++++), C(++--), A(+--+) and G(+-+-) for Ru2 and Ru3.

| IrRep | Ru1 and Ru4 | | | Ru2 and Ru3 | | | Space group |
|---|---|---|---|---|---|---|---|
| P*bam* | X | Y | z | x | y | z | |
| $\Gamma_1$ | -- | -- | $a_z$ | -- | -- | $G_z$ | P*bam* |
| $\Gamma_2$ | -- | -- | $f_z$ | -- | -- | $F_z$ | P*b'a'm* |
| $\Gamma_3$ | $f_x$ | $a_y$ | -- | $F_x$ | $G_y$ | -- | P*b'am'* |
| $\Gamma_4$ | $a_x$ | $f_y$ | | $G_x$ | $F_y$ | -- | P*b'am'* |
| $\Gamma_5$ | | | | -- | -- | $A_z$ | P*b'a'm'* |
| $\Gamma_6$ | | | | -- | -- | $C_z$ | P*bam'* |
| $\Gamma_7$ | | | | $C_x$ | $A_y$ | -- | P*b'am* |
| $\Gamma_8$ | | | | $A_x$ | $C_y$ | -- | P*b'am* |

We argue that an antiferromagnetic ordering of the type $A_xC_y$ (or $C_xA_y$) on exclusively the outer ruthenium atoms (corresponding to the irreducible representation $\Gamma_8/\Gamma_7$ ) can account for the large difference in the intensity of the two reflections. In the irreducible representation $\Gamma_8$ the inner ruthenium atoms do not carry a magnetic moment whereas the outer ruthenium atoms have a $A_xC_y$ ordering, i.e. an A mode along x, a C mode along y and zero moment along z: $+M_x,+M_y, 0$; $-M_x,+M_y$ 0; $-M_x,-M_y, 0$; $+M_x,-M_y, 0$. The square magnetic structure factor for the (00L) reflections, (assuming that the ruthenium atoms at $z_1$ and $z_2$ be in phase) would write:

$$[F(00l)]^2 = 16M_{out}^2 \cos^2[l\pi(z_1 + z_2)]\sin^2[l\pi(z_1 - z_2)] \sim 16M_{out}^2 \sin^2[l\pi 0.22]$$

In this model the only magnetic reflection having a sizable intensity is the 002, which is indeed what we observe experimentally (Fig. 2).

We can estimate the value of $M_{x/y}$ being approximately half Bohr magneton from the ratio of the magnetic to the nuclear contribution $I_M/I_N$ for the 002 reflection.

Table T3 *Intensities in arbitrary units of the measured magnetic reflections in $Sr_4Ru_4O_{10}$ as calculated with the program Fullprof [24] for the two models corresponding to the irreducible representations $\Gamma_2$ and $\Gamma_8$. As already reported [14], the IrRep $\Gamma_8$ does not allow a magnetic moment on the inner ruthenium atoms. Ru1 and Ru4 (Ru_in) but only on the outer ruthenium atoms, Ru2 and Ru3 (Ru_out). The magnetic intensity in the 00L reflections is only observed on the 002.*

| I(a.u.)/$\Gamma$ | Ru1 Ru4 | Ru2 Ru3 | 002 | 004 | 006 | 008 | 0010 | 0014 |
|---|---|---|---|---|---|---|---|---|
| $\Gamma_3(\Gamma_4)$ | $f_x$ $a_y$ - | $F_x$ $A_y$ - | 13 | 10 | 0.14 | 0.96 | 0.47 | 0.1 |
| $\Gamma_8(\Gamma_7)$ | --- | $C_x$ $A_y$ - | 15 | 0.46 | 0.81 | 0.21 | 0.06 | 0 |
| **Iobs (a.u.)** | | | 15 ±2 | -- | -- | -- | -- | -- |

We have mainly focussed on the 00L reflections because they are sensitive only to the components of the magnetic moments in the *ab* plane and do not probe the moment along z. As a consequence, the additional intensity at T* adds to a signal which is purely nuclear. The fact that the additional intensity peaks at T* and almost vanishes as the temperature is lowered to T=1.5 K would seem to rule out the possibility of an ordered canted structure [21] (ferromagnetic modes along z but antiferromagnetically canted in the *ab* plane) at T=2K.

It is worth stressing that in the case of a magnetic structure with propagation vector K=(0,0,0) (which encompasses the ferromagnetic order and some antiferromagnetic arrangements), the magnetic intensities do not appear as separate peaks in the reciprocal space because the magnetic cell is equal to the nuclear cell. The magnetic scattering manifests itself as additional intensity on top of the nuclear reflections in the temperature range where the order sets in but no scattering is observed between nuclear nodes. In our case, the antiferromagnetic contribution appears at about 50K on existing peaks and, within our experimental conditions, is not present at T=2K. At T=2K the intensity is different from zero because of the nuclear contribution and it is indeed comparable to what observed at T=100K (Fig.2).

Conversely, the temperature dependence of the H H 0 reflections (see Figure 3) shows a clear dip at around T*, indicating that the total $M_z$ component of the magnetic moments is sensibly reduced at the temperature where the $M_{xy}$ component is maximised. In Table T3 we compare the intensities of the main magnetic reflections as calculated with the program Fullprof [21] in the case of the only two candidate irreducible representations that contain in plane magnetism: the $\Gamma_3$ which involves also ferromagnetism in the *ab* plane and the $\Gamma_8$ which has antiferromagnetism in the *ab* plane but does not allow any contribution from the inner ruthenium atoms. All the other irreducible representation had to be discarded since they yield calculated intensities I(HKL) at odds with the experimental observations, namely the $\Gamma_5$ and $\Gamma_6$ would yield vanishing intensities for all the 00L reflections.

As prevoulsy reported [17, 18], in $Sr_4Ru_3O_{10}$ there is a clear correlation between magnetic and structural effects, in particular the metamagnetic transition occurs at a temperature T*=50K for which the cell volume has a sharp minimum. Whilst the *a* paramater has a monotonic dependence on temperature, the *c*-axis decreases down to 50K [17] and is clearly much more affected by temperature and or magnetism. Theoretical calculations have pointed out that in the regime of *c*-axis elongated, the spin–orbit coupling in the $RuO_6$ octahedra would tend to favour magnetic correlations along the *c*-direction, while for in-plane elongated octahedra the orbital occupation would cooperate with the spin–orbit for inducing local spin moments in the *ab*-plane [25,26].

On the basis of these considerations, one can envisage a scenario in which the two substructures of inner and outer ruthenium atoms act partly as independent players in determining the magnetic order in $Sr_4Ru_3O_{10}$. This would imply that the hamiltonian describing the system contains terms of order higher than 2 [27]. At $T_c$=105K the substructure of the inner ruthenium atoms orders ferromagnetically along z whilst the outer ruthenium atoms would tend to align antiferromagnetically in the *ab* plane. This *ab* antiferromagnetic ordering is partly achieved at T* where the *c* axis has a minimum [17] and the anisotropy along *c* is reduced. At around T* the irreducible representation $\Gamma_2$ (ferromagnetism along z with $F_z \neq 0$ and $f_z \sim 0$) would coexist with the $\Gamma_8$ describing an antiferromagnetic order of the type $A_xC_y$ uniquely in the substructure of the outer ruthenium atoms. In other words, below $T_C$ the antiferromagnetic and ferromagnetic interactions are in competition, with the inner ruthenium atoms more constraint along z and the outer ruthenium atoms more free to bend away from the easy axis. As the temperature is lowered to 2K, the *c* axis becomes longer and the stronger anisotropy along z forces all ruthenium atoms to line up along *c*. The alignement of the outer ruthenium atoms however may be incomplete as suggested not only by the smaller value of $M^z_{out}$ at 2K but also by the temperature dependence of the reflections. In Figure 3 we draw the 220, 22-16 and 111 reflections. They all show a dip corresponding to the metamagnetic transition at 50K (where the 002 increases) but they behave differently on lowering the temperature to 2K. Namely, the H,H,L reflections increase significantly more than the 220 since the HHL probe the components slightly off axis: the 111 reflection is sensitive to the moments lying in the (111) plane which is at about 18° from the *c*-axis.

It is also instructive to examine the behaviour of some key reflections in the presence of a magnetic field applied in the *ab* plane when unpolarised neutrons are diffracted: the 008 and the 110 reflections for instance have a rather small nuclear contribution and probe components of the magnetic moment in the (HK0) plane and in the (HHL) plane respectively. At T=2K, both reflections show vanishing hysteresis but have a rather different behaviour: the 008 increases significantly whilst the 110 stays almost constant (Figure 4).

The field tends to orient the disordered off-axis moments of the outer ruthenium atoms, yielding a component *ab* whitout reducing significantly the total $M^z$. Conversely, at 70K just above the metamagnetic transition T*, the intensity of the HH0 reflections is sensibly reduced with the field (Figure 5) because in the proximity of the metamagnetic transition the moments on the outer ruthenium atoms are softer and can be easily bent away from the *c*-axis into the *ab*-plane.

Such a scenario is further supported by polarised neutron measurements [25] conducted with an external field applied in the *ab* plane. In the magnetisation maps [25], the outer ruthenium atoms appear to be sensibly more magnetised at 65K than at 2K whereas the inner ruthenium atoms stayed unchanged throughout the metamagnetic temperature T*. The 004 reflection is again a good test of what happens at T*: unlike the other reflections, it goes from a negative to a positive value throughout T* [25] which, in a regime of in plane polarisation, indicate a stronger anisotropy along *c* for T<T*.

On the basis of a single observation is not possible to solve quantitatively the magnetic structure at T*, but we have nonetheless some indication that at the metamagnetic transition an antiferromagnetic order appears in the *ab* plane being triggered either by temperature or some other effect. The "loss" of magnetic moment which has been observed recently [28] with magnetization measurements as a function of a magnetic field rotating between H $\perp$ c and H $\parallel$ c might be explained in terms of this partial AFM alignment.

Antiferromagnetic correlations at intermediate temperatures are not unusual in ruthenates. As an example, in the bilayered compound $Sr_3Ru_2O_7$ magnetic fluctuations have been detected and shown to evolve from a ferromagnetic position to an incommensurate antiferromagnetic vector as the temperature decreases from 115K to 15 K [29]. Similarly in the single layer $Sr_2RuO_4$, antiferromagnetic fluctuations were likewise observed at low temperature at an incommensurate position in the [hh0] direction [30]. In the triple layer $Sr_4Ru_3O_{10}$ the anomaly observed in the specific heat at $T_c$=105K [15] is an order of magnitude smaller than expected for a complete spin ordering corroborating a picture in which only the inner ruthenium atoms contribute initially to the ferromagnetic transition whereas the outer ones tend to be antiferromagnetically ordered. There exists therefore a delicate balance between ferro and antiferromagnetic interactions in ruthenates, and this balance can be altered by temperature or external pressure. In $Sr_4Ru_3O_{10}$ the application of a modest hydrostatic pressure for instance diminishes the *c*-axis ferromagnetism and induces basal plane antiferromagnetism as recently shown with resistivity measurements [31].

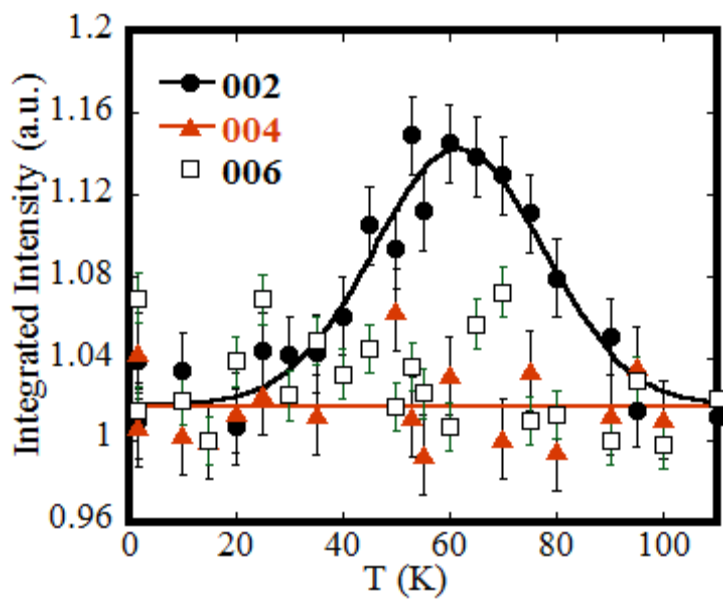

Figure 2 The neutron intensity of the 002, 004, 006 Bragg reflections as a function of temperature, normalized to their respective values at T=115K. The curves were measured on the diffractometer D10 at ILL.

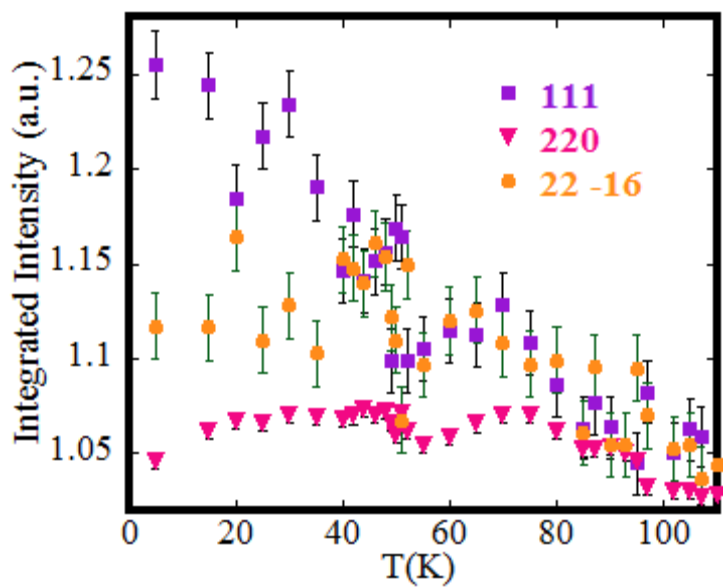

Figure 3 The neutron intensity of the 111, 220, 22-16 Bragg reflections as a function of temperature, normalized to their respective values at T=115K. The curves were measured on the diffractometer D10 at ILL.

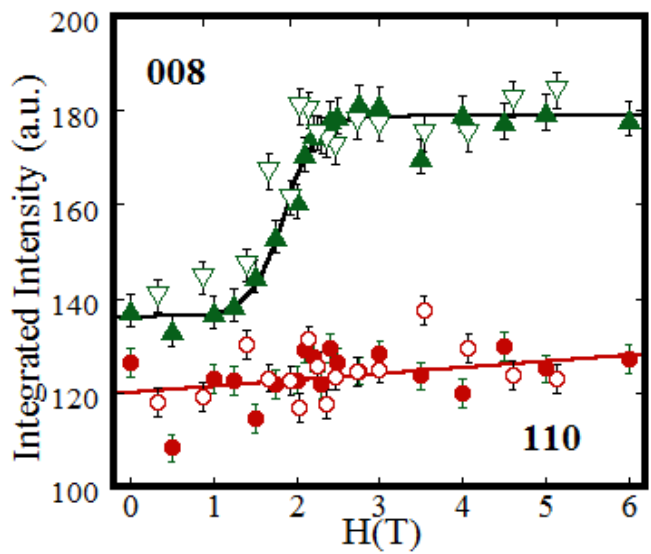

Figure 4 The intensity of the Bragg reflections 008 and 110 as a function of a magnetic field, in the *ab* plane as measured on the D10 diffractometer at ILL at 1.5 K. The full and empty symbols refer to the up and down sweeps.

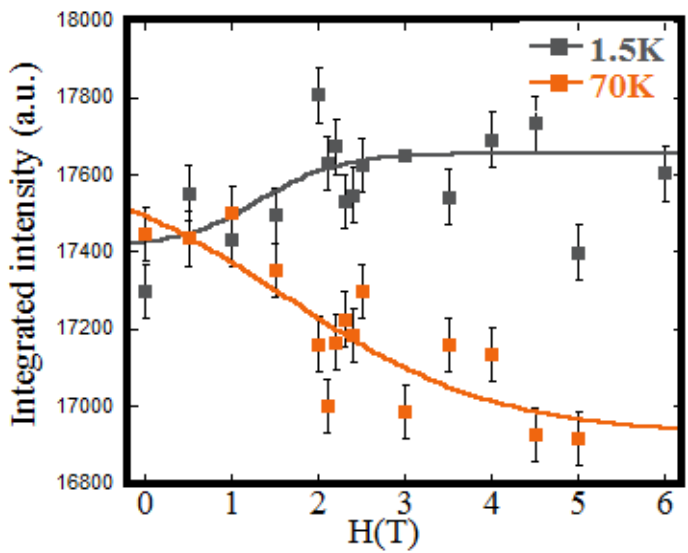

Figure 5 The intensity of the Bragg reflections 220 as a function of an *ab* applied magnetic field, as measured on the D10 diffractometer at ILL. The upper curve was taken below the metamagnetic transition at T* and the lower curve above T*.

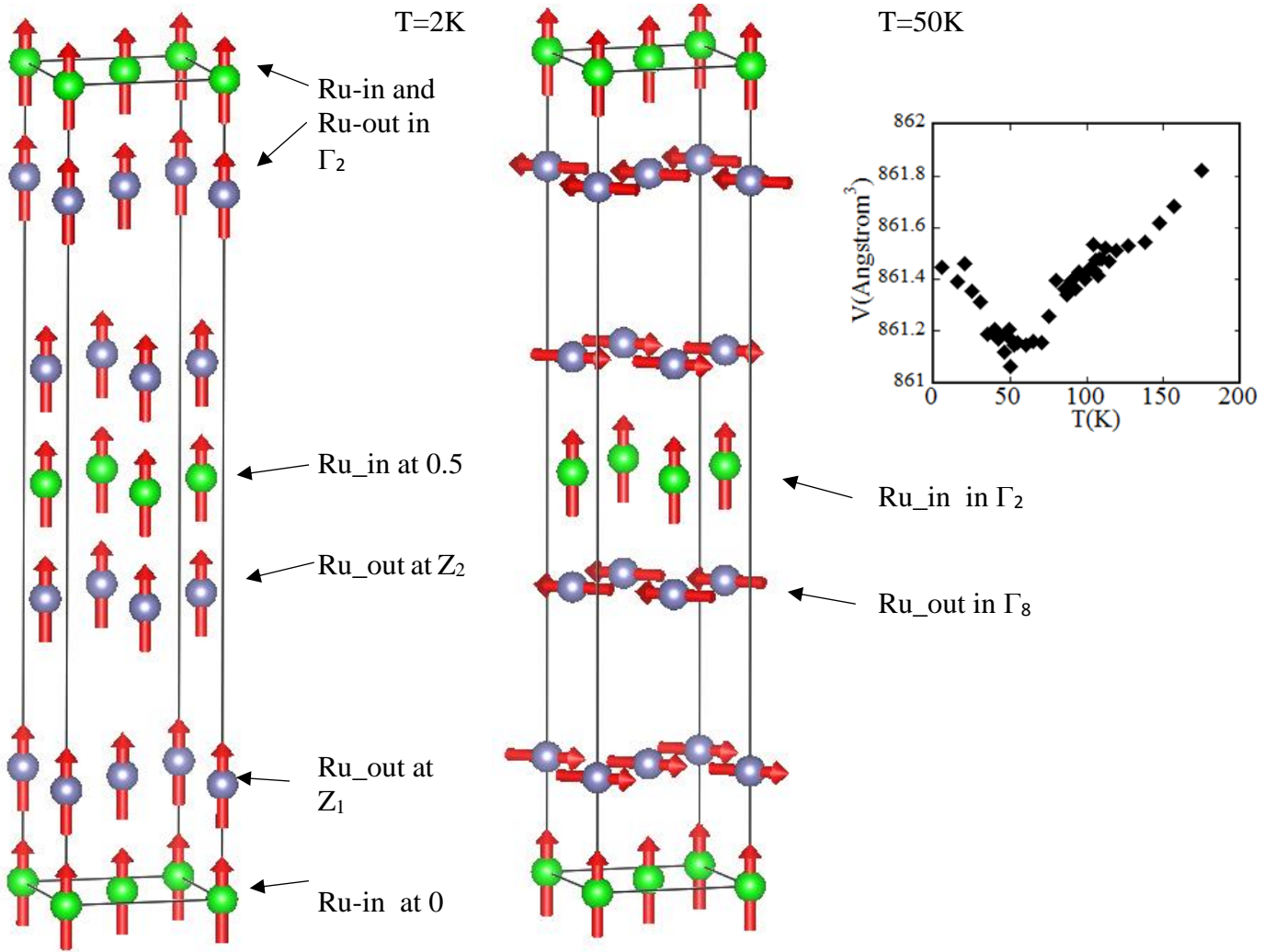

Figure 6a Magnetic structure at T=2K [17]. Both substructures of ruthenium atoms, inner and outer, are in the $\Gamma_2$ irreducible representations for the magnetic modes. The $c$ axis is the long one.

Figure 6b Proposed magnetic structure at T*=50K. The substructures of inner ruthenium atoms is in the $\Gamma_2$ irreducible representations. The outer ruthenium atoms are in the $\Gamma_8$. In the inset, the cell volume as a function of temperature [17] shows a sharp minimum at the metamagnetic transition at T*=50K.

**Conclusions**

Using neutron diffraction we have made a further step in understanding the origin of the metamagnetic transition in the triple-layer ruthenate $Sr_4Ru_3O_{10}$. The two substructures consisting of the inner and outer ruthenium atoms of the trilayers seem to have a different temperature evolution around the metamagnetic temperature, in particular the antiferromagnetism appearing about T* seems to be driven only by the outer ruthenium atoms with the trilayers at the face centre in phase with the blocks located in the basal plane. The substructure of the inner ruthenium atoms does not contribute to the antiferromagnetic order but are ferromagnetically ordered along the $c$-axis.

**Acknowledgements**

We would like to thank Oscar Fabelo and Anne Stunault for their assistance and help during the experiments and Mario Cuoco and Filomena Forte for useful discussions.

**References**

1. Maeno, Y. *et al Journal of the Physical Society of Japan* **66**, 1405–1408 (1997).

2. Ikeda, S.-I., *et al, Phys. Rev. B* **62**, R6089–R6092 (2000).

3. Kanbayasi, A. *Journal of the Phys.Soc. of Japan* **41**, 1879–1883 (1976).

4. Allen, P. B. *et al, Phys. Rev. B* **53**, 4393–4398 (1996).

5. Nakatsuji, S., *et al, Journal of the Phys.Soc. of Japan* **66**, 1868 (1997).

6. Cao, G. *et al, Phys. Rev. Lett.* **78**, 1751–1754 (1997).

7. Yoshida, Y. *et al, Phys. Rev. B* **69**, 220411 (2004).

8. Shepard, M. *et al, Journ. of App. Phys.* **79**, 4821–4823 (1996).

9. Cao, G., *et al, Journ. of App. Phys.* **81**, 3884–3886 (1997).

10. Crawford, M. K. *et al, Phys. Rev. B* **65**, 214412 (2002).

11. Cao, G. *et al, Phys. Rev. B* **68**, 174409 (2003).

12. Carleschi E. *et al*, *Phys. Rev.* B **90**, 205120 (2014).

13. Weickert F. *et al*. *Physica B* **536,** 634 (2018).

14. Yan Liu *et al. New J. Phys.* **18,** 053019 (2016).

15. Lin, X. N. *et al*, *Solid State Communications* **130**, 151–154 (2004).

16. Bao W. *et al*, arXiv :cond-mat/0607428v1(2006).

17. Granata V. *et al, J. of Phys. : Cond. Matter* **25**, 056004 (2013).

18. Gupta R. *et al, Phys. Rev. Lett* **96**, 067004 (2006).

19. Yan Liu . *et al*, *Phys. Rev.* B **98**, 024425 (2018).

20. Zhu M. *et al, Scient. Rep.* **8,** 3914 (2018).

21. Fittipaldi R. *et al*, Crystal Growth and Design **7**, 2495 (2007).

22. Singh D.J. et Mazin I. I. *Phys. Rev. B* **63**, 165101(2001).

23. Capogna L. *et al*, Applied Physics A, **74** I S926-S928 (2002).

24. Rodriguez Carvajal J. *Physica B*. **192**, 55 (1993).

25. Granata V. *et al, Phys. Rev. B* **93**, 115128 (2016), Forte F. et al, arXiv :1812.01359v2 (2019).

26. Cuoco M, Forte F and Noce C. *Phys. Rev.* B **73**, 094428 (2006).

27. Bertaut E.F. Acta Cryst. **A 24,** 217 (1968)

28. Weickert F. *et al, Scient. Rep.* **7,** 3867 (2017).

29. Capogna L. *et al,* P*hys. Rev. B* **67**, 012504 (2003).

30. Sidis Y *et al, Phys. Rev. B* **83**, 3320 (1999).

31. Zheng H. *et al, Phys. Rev. B* 98, 064418 (2018).